\documentclass [11pt]{article}
\usepackage[dvips]{color}
\usepackage{epsfig}
\usepackage[normalem]{ulem}
\def\beq{\begin{equation}}
\def\eeq{\end{equation}}
\def\beqa{\begin{eqnarray}}
\def\eeqa{\end{eqnarray}}
\def\d{{\rm d}}

\begin{document}
\baselineskip0.6cm plus 1pt minus 1pt
\tolerance=1500

\begin{center}
{\LARGE\bf 
Thermodynamical asymmetries in \\
whirling, jumping and walking}
\vskip0.4cm
{ J. G\"u\'emez$^{a,}$\footnote{guemezj@unican.es},
M. Fiolhais$^{b,}$\footnote{tmanuel@teor.fis.uc.pt}
}
\vskip0.1cm
{\it $^a$ Departamento de F\'{\i}sica Aplicada}\\ {\it Universidad de
Cantabria} \\ {\it E-39005 Santander, Spain} \\
\vskip0.1cm
{\it $^b$ Departamento de F\'\i sica and Centro de
F\'\i sica Computacional}
\\ {\it Universidade de Coimbra}
\\ {\it P-3004-516 Coimbra, Portugal}
\end{center}

\begin{abstract}
We analyze, from the thermodynamical point of view, mechanical systems in which there is production of mechanical energy due to an
internal source of energy, and compare that analysis with the similar one for the ``symmetric" motion which occurs with energy dissipation.
The analysis of the energetic asymmetries is instructive to put in evidence the role of thermodynamics even in the discussion of mechanical aspects.
We illustrate the discussion with the well known example of a person on a rotating platform outstretching and contracting his/her arms, and also with other common situations such as jumping and walking.

\end{abstract}

\section{Introduction}
\label{sec:intro}

Classical mechanics is time reversal invariant, i.e. by changing $t\rightarrow -t$ the fundamental Newton equation (second law of mechanics)
remains invariant. However, the second law of thermodynamics, that applies to all systems, including those that are usually categorized as
mechanical ones, clearly indicates the ``arrow of time", imposing a direction for their evolutions. Our focus in this paper is on the discussion of certain pendular motions, containing parts that are symmetric,
sometimes in the time reversal sense, but that are asymmetric thermodynamically speaking.

If there were no dissipation of mechanical energy, a particle, such as a simple pendulum, would remain in motion forever. Clearly, this cannot be the case in real systems. However, real systems, even if they are dissipative, may perform symmetric (pendular) motions but that requires an internal (or external) source of energy. If that source gets exhausted, the system eventually stops.

In this paper we discuss several mechanical systems --- a whirling person, a vertical jump and a steady state walk ---,  emphasizing the role played by the first law of thermodynamics
in addition to the Newton's second law. Both are fundamental and independent laws of nature, therefore they are always valid for any system undergoing whatever process.

For a system of {\em constant} mass $M$, the Newton's second law states that
the resultant of the external forces, ${\vec F}_{\rm ext}= \sum_j {\vec F}_{{\rm ext}, j}$, is equal to that mass times the centre-of-mass acceleration, or equivalently
$
 M \d {\vec v}_{\rm cm}={\vec F}_{\rm ext}\  \d t  \ ,
$
where $\vec v_{\rm cm}$  stands for the  centre-of-mass velocity.
The integral form of that equation is $\Delta \vec p_{\rm cm}= \vec I $, where
$ \vec I = \int  {\vec F}_{\rm ext}\  \d t $ is the impulse of the resultant of the external forces, and states that this external impulse,  in a given time interval, is equal to the
variation, in the same time interval, of the system centre-of-mass linear momentum, $\vec p_{\rm cm}=M \vec v_{\rm cm}$.

Newton's fundamental equation can be expressed in other equivalent forms, e.g. it also states that
the infinitesimal variation of the centre-of-mass kinetic energy equals the
``pseudo-work" \cite{penchina78,sherwood83},
which is the dot product of the external resultant force  by the infinitesimal displacement of the
centre-of-mass: $
 {1\over 2} M \d v_{\rm cm}^2={\vec F}_{\rm ext} \cdot \d {\vec r}_{\rm cm}\, ,
$
or
\beq
\d K_{\rm cm} = {\vec F}_{\rm ext} \cdot \d {\vec r}_{\rm cm} \, , \label{newtonpt2}
\eeq
where $K_{\rm cm}={1 \over 2} M v^2_{\rm cm}$ is the centre-of-mass kinetic energy.
The integral form of  equation (\ref{newtonpt2})  is
$\Delta K_{\rm cm}={W_{\rm ps}}$, where the left-hand side is the variation of the centre-of-mass kinetic energy
and the right-hand side is the ``pseudo-work" of the resultant external force, ${W_{\rm ps}}=\int {\vec F}_{\rm ext} \cdot \d {\vec r}_{\rm cm}$ \cite{mallin92}.
It is worthwhile to make  clear the distinction between work and pseudo-work: in the former, one considers the displacement of the force itself, whereas in the
latter, corresponding to the present case,  it is the displacement of the centre-of-mass that matters. Hence, the energy-like formulation of Newton's second law (kinetic energy theorem) uses rather the pseudo-work \cite{mungan05} and not the real work, and this is a point that, surprisingly, is not much emphasized in textbooks.

On the other hand, the first law of thermodynamics is a statement on energy conservation and,
for a general infinitesimal (reversible or irreversible) process on a system \cite{jewett08v},
it can be expressed as \cite{besson01,erlichson84}
\beq
\d {K}_{\rm cm} + \d { U} = \sum_j { {\vec F}_{{\rm ext}, j}}\cdot \ \d {\vec r}_j  \ + \ \delta { Q}.
\label{totale1}
\eeq
where $U$ is the internal energy and
 $\delta Q$ is the infinitesimal heat which  is not an exact differential, contrary to $\d U$, $\d K_{\rm cm}$ or $\d {\vec r}_j$ that are scalar or vector exact differentials.
Each term in the sum over $j$ on the right-hand side of equation~(\ref{totale1})
is a work associated with each {\em external} force ${\vec F}_{{\rm ext}, j}$, and  $\d {\vec r}_j$
is the corresponding infinitesimal displacement of that  force itself (and not, anymore, the displacement of the centre-of-mass). Therefore, $\delta  {W}_j= {\vec F}_{{\rm ext}, j} \cdot \d {\vec r}_j$
is real work (and not pseudo-work).
The integral form of equation (\ref{totale1}) is expressed by $\Delta K_{\rm cm}+ \Delta U= W_{\rm ext} + Q$.
In expressing the first law of thermodynamics in this way, one assumes that any translational kinetic energy with respect to the centre-of-mass and the rotational energy should be included in the internal energy of the system, a point that will be illustrated with the example of the next section \cite{guemez13,guemez14}.

In the next sections we study several examples involving the motion of a person, with the common feature that there is always a need for a source of internal energy. In humans and, in general, in animals, this energy is provided by biochemical reactions that take place inside the body. These are usually complicated and irreversible processes but, as explained in the next section, by assuming a simple model with reversible processes
and with the Gibbs free energy playing a central role, we are able to pinpoint some interesting energetic issues.

\section{Rotating person}

\label{sec:homem}

In many university physics textbooks a person on top of a turntable, freely rotating around a vertical axis,
contracting and outstretching the  arms, is presented  \cite{walker11c}
as an example of a zero external torque system, therefore useful to illustrate the angular momentum conservation \cite{tipler04}.
However the associated energetic issues, such as the production or the dissipation of rotational kinetic energy, when the arms are pushed towards the body or outstretched, are usually not considered or expressed in equations  \cite{rees94}.
At first sight, the two processes --- arms contraction and outstretching --- may seem symmetric in many respects but, from the energetic point of view, they are not.

\begin{figure}[htb]
\begin{center}
\hspace*{-0.5cm}
\includegraphics[width=9cm]{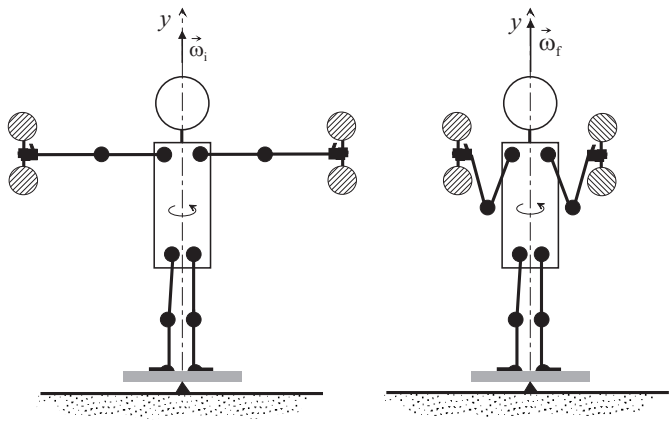}
\end{center}
\vspace*{-0.5cm}
\caption[]{\label{fig:homem} \small A person  on top of a rotating turntable, initially with outstretched arms, moves them towards the body. The dumbbells just enhance the
difference between the initial and final moments of inertia (and, therefore, the difference in magnitude between the initial and final angular velocities). }
\end{figure}

In figure~\ref{fig:homem} a person holding a pair of dumbbells on top of a rotating turntable is represented, first with the arms outstretched and, finally, with the arms close to the body.
A similar example is provided by a whirling ice skater, or a whirling ballet dancer, etc.
The initial angular velocity is $\omega_{\rm i}$ and it changes, due to the conservation of angular momentum, when the person's arms position and, therefore, the moment of inertia, changes. We denote by $\omega_{\rm f}$ the final angular velocity and by $I_{\rm i}$ and by $I_{\rm f}$ the initial and final moments of inertia, respectively.

The external forces are the normal reaction and the weight, both producing a vanishing resultant force and a vanishing external torque, ${\Gamma}_{\rm ext}=0 $. For a variable moment of inertia rotating system
the Newton's second law is better expressed by  $ {\Gamma}_{\rm ext}\d t = \d (I\omega)$. Therefore,
in the course of the process, both the angular velocity, $\omega$, and the moment of inertia, $I$, change, but the angular momentum, $L$, remains constant:
\beq
L= I \omega= I_{\rm i} \omega_{\rm i} = I_{\rm f} \omega_{\rm f} \, .
\label{momang}
\eeq

Let us now consider the energetic point of view as given by the first law. In order to contract the arms, biochemical reactions, that generically we denote by $\xi$, should take place in the person's body \cite{atkins10}. These reactions cause variations of internal energy, $\Delta U_\xi$, volume, $\Delta V_\xi$, entropy, $\Delta S_\xi$, etc.
If a chemical reaction takes place at constant external pressure, $P$, in diathermic contact with a heat reservoir at temperature $T$, part of the internal energy is used for an expansion against the external pressure,$W_\xi=- P \Delta V_\xi$,
 and part must be exchanged with the heat reservoir in order to ensure that the entropy of the universe does not decrease, $Q_\xi =  T \Delta S_\xi $.
When $\Delta V_\xi < 0$, the external pressure performs work on the system and when $\Delta S_\xi > 0$ the heat reservoir
increases the internal energy of the body. We assume such a simple model that no other  energy is exchanged between the person and the environment.

On the left-hand side of the integral form of equation~(\ref{totale1}), $\Delta K_{\rm cm}=0$ and the contributions to
the total internal energy variation ($\Delta U$) just come from the variation of the rotational kinetic energy ($\Delta U_{\rm R}$) and from the biochemical reactions, that take place  in the person's
body ($\Delta U_{\xi}$): $\Delta U = \Delta U_{\rm R} + \Delta U_{\xi}$.  Regarding $\Delta U_{\rm R}$, it is given by
\beq
\label{eq:rkeerp}
\Delta U_{\rm R} = {1\over 2} I_{\rm f} \omega_{\rm f}^2 - {1\over 2} I_{\rm i} \omega_{\rm i}^2\, ,
\eeq
or, using (\ref{momang}), by \cite{chabay11}
\beq
\Delta U_{\rm R} = {L^2\over 2}  \left({1 \over I_{\rm f}} -   {1 \over I_{\rm i}}\right)\, .
\label{eq:rkeerpx}
\eeq
Since $L$ is constant, if $I$ decreases in the process, as in figure \ref{fig:homem}, then $\Delta U_{\rm R}$ increases, and vice-versa (figure \ref{fig:homem}, from right to left).

We first consider $I_{\rm f}<I_{\rm i}$ as described by figure \ref{fig:homem}.
The weight and the normal force do not perform any work, so that the only contributions to the right-hand side of the integral form of equation (\ref{totale1})
are $W_P= - P \Delta V_\xi$ and $Q=T\Delta S_\xi$, the thermodynamical work and the heat associated with the chemical reactions that take place in the
  person's body  \cite{guemez13}. Both $\Delta V_\xi$ and $\Delta S_\xi$ (referring to the system) are equal in magnitude to the same
  variations experienced by the atmosphere that behaves as a heat and a work reservoir. Therefore, the integrated equation (\ref{totale1}) leads to
\beq
{1\over 2} I_{\rm f} \omega^2_{\rm f}- {1\over 2} I_{\rm i} \omega^2_{\rm i} + \Delta U_\xi = - P \Delta V_\xi + T \Delta S_\xi
\eeq
or
\beq
{1\over 2} I_{\rm f} \omega^2_{\rm f}- {1\over 2} I_{\rm i} \omega^2_{\rm i}= - \Delta G_\xi
\label{arms1}
\eeq
where $\Delta G_\xi= \Delta U_\xi+P\Delta V_\xi - T\Delta S_\xi$ is the  Gibbs free energy variation in the person's body. The increase of kinetic energy results, after all, from the corresponding expend of energy by the person to pull the dumbbells closer to the body.  Hence, when the person moves the arms inward towards the  body, there is a decrease of the  Gibbs free energy, and the process is assumed to be,
at least in principle,  reversible. Actually, the increment of  rotational energy can be subsequently be transformed into another kind of energy and, in particular, it may lead to a Gibbs free energy increment, $-\Delta G_\xi$, of a certain chemical reaction elsewhere. In an even more simplified (but still reliable) model we could ignore the (reversible) work and the (reversible) heat exchanged with the atmosphere, only considering the
(reversible) variation of the internal energy. This would become similar to the case when the source of internal energy is of a mechanical nature, such as the energy stored by a spring responsible for the motion of a toy \cite{guemez09}.

However, when $I_{\rm f} > I_{\rm i}$ (for example, when the person outstretches the arms), the
rotational energy decreases and there is no way to transform the dissipated mechanical energy
into an increment of Gibbs free energy. In this case  one has simply $\Delta U = \Delta U_{\rm R}$. Spontaneously, this  mechanical energy decrease dissipates as energy transferred  (heat) to the  surroundings.
Therefore, the first law of thermodynamics leads to an equation that differs from (\ref{arms1}).  Indeed, the relevant energy equation for the arm outstretching process is
given by
\beq
{1\over 2}I'_{\rm f} {\omega'}_{\rm f}^2 - {1\over 2}I'_{\rm i} {\omega'}_{\rm i}^2   = Q\,
\eeq
or, using (\ref{eq:rkeerp}) and (\ref{eq:rkeerpx}),
\beq
Q = {L^2\over 2}  \left({1 \over I'_{\rm f}} -   {1 \over I'_{\rm i}}\right)\, ,
\eeq
which is a negative quantity because $I$ increases in the process.
When the moment of inertia increases
there is an increment of the entropy of the universe given by
$
\Delta S_{\rm U} = - {Q\over T} > 0\, .
$

Although in both processes -- outstretching the arms or pulling them towards the body -- the
angular momentum is conserved, the rotational kinetic energy varies. From the  energetic point of view, the processes are totally
different, an important  point that is not tackled in textbooks.
When the arms are contracting, it is acceptable to assume an ideal reversible process to describe the situation. In the final state of this phase of the process, that proceeds with a reduction of the Gibbs function, the initial energy is still stored as kinetic energy of the system, i.e. as ``organized" energy (the real process, contrary to this idealized one, is irreversible, so a part of the variation of the Gibbs function is dissipated energy
and there is an associated increase of the entropy of the universe).
In the other phase of the process, when the rotating person outstretches the arms, there is no internal mechanism for compensating the decrease of mechanical (kinetic) energy through an increase of chemical energy and the kinetic energy variation is inevitably dissipated (heat) in the surroundings.
The entropy clearly increases and the process is definitely  irreversible.
After a complete cycle (pulling and outstretching the arms), the net result is a decrease of the Gibbs energy in the person's body (first part of the process) that goes to heat dissipated in the surroundings (second part of the process).
Therefore a person cannot
remain indefinitely in rotation, contracting and outstretching the arms in a kind of permanent pendular motion! Just to keep the exercise, in addition to the basal metabolism, after some time he/she has to eat something...

One may take a different (microscopic) perspective to discuss the encountered thermodynamic asymmetry, namely based on
different numbers of degrees of freedom. More precisely, when the person pushes the arms towards the body, there is a, so to say, flux of inner microscopic energy (large number of degrees of freedom) towards the person's body (small number of degrees of freedom). Because of the concentration of the energy in a small number of degrees of freedom, it is likely that the process can be made almost without energy dissipation. On the opposite case (the person outstretching the arms), the energy would have to go from the persons's body (small number of degrees of freedom) to the large number of degrees of freedom inside the body. It is then much more difficult to control all the energy spreading in many pieces, and the process is indeed generally irreversible \cite{leff96}. In the real situation, both phases of the process are irreversible, but the second one is `more irreversible'
(more increment of the entropy of the universe) than the other one.

As in \cite{guemez13,guemez14}, here we also would like to emphasize that the total kinetic energy of a composite, articulated body may change, even when no work is done by external forces.
In the description of processes with production of mechanical energy, such as the present example, when $I_{\rm f}<I_{\rm i}$ the chemical reactions
play a central role.

The
asymmetry encountered in this energetic discussion is of the same nature of the asymmetry that will also be found in the examples presented in the
next sections.

\section{Jumping person}

\begin{figure}[htb]
\begin{center}
\hspace*{-0.5cm}
\includegraphics[width=13cm]{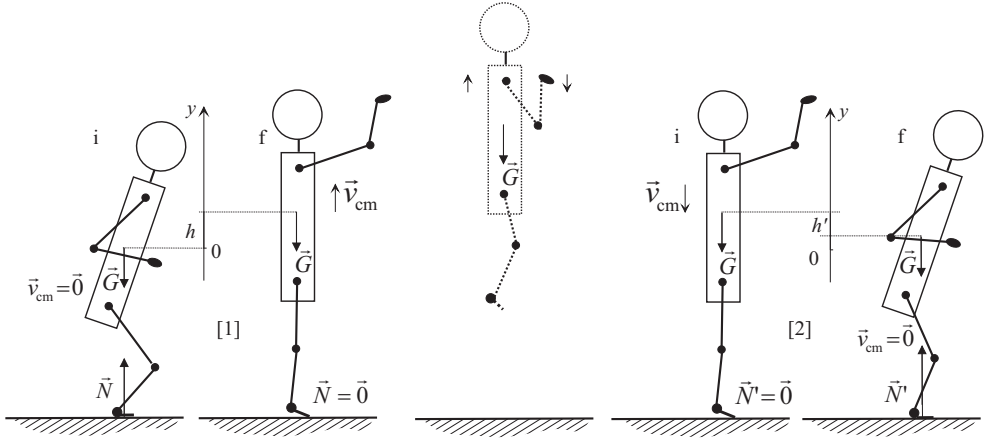}
\end{center}
\vspace*{-0.5cm}
\caption[]{\label{fig:jumper} \small A person jumping vertically (1) still contacting the ground in the elevation phase (2) again in contact with the ground in the descending; i and f refer to the initial and final states of each phase. The magnitude of the centre-of-mass velocity of the final state of phase 1 is the same as the magnitude of the initial state  of phase 2.  The intermediate phase (middle picture) is not important for the discussion. }
\end{figure}

Let us consider the motion of a person of mass $M$ jumping vertically. We are interested in the part of the jump during which there is still contact of the foot with the ground. This example was already presented in \cite{guemez13d} in a different context, but it is useful to
revisit it here in the context of the present thermodynamical discussion.
We assume the following simplifying assumptions: the centre-of-mass motion is vertical, and the normal force, $\vec N$, is constant. In this framework, the problem reduces to one dimension and the integration of equation (\ref{newtonpt2}) is straightforward. If the person raises its centre-of-mass by $h$ (see figure \ref{fig:jumper}, part 1), that equation leads to
\beq
{1\over 2} M v_{\rm cm}^2 + Mgh= N h\, ,
\label{yudfg}
\eeq
where $v_{\rm cm}$ is the centre-of-mass speed when the feet are loosing contact with the ground. Hence,  the increase of mechanical energy in the process is equal to the pseudo-work of the normal force.

That normal force is the reaction to the force exerted by the person on the ground and, to produce it, biochemical reactions should take place in the person's muscles as discussed in the previous section. Under the same simplified assumption considered in Sec. 2 (chemical reactions at constant pressure and temperature), and applying the arguments exposed therein,
the first law of thermodynamics for this initial phase of the jump, during which there is still contact with the ground, leads to
\beq
{1\over 2} M v_{\rm cm}^2 + \Delta U_\xi = -Mgh -P \Delta V_\xi + T\Delta S_\xi\, .
\label{eq:fltboys}
\eeq
or, equivalently
\beq
{1\over 2} M v_{\rm cm}^2 +Mgh = - \Delta G_\xi\, ,
\label{anter}
\eeq
where the Gibbs free energy variation turns out to be the symmetric of the pseudo-work of force $\vec N$ [see equations (\ref{yudfg}) and (\ref{anter})]. As equation (\ref{anter})
shows,  the person's increase of centre-of-mass kinetic and potential energy is ultimately due to biochemical reactions \cite{atkins10}.
The variation of the Gibbs free energy is the maximum useful work that can be obtained from the reactions in the  person's muscles biochemical reactions: $W_{\rm max} = -  \Delta G_\xi$ and this is also equal to $Nh$.
We stress that the normal force does not do any work but it intermediates an energy transformation.

The initial phase of  this jump is a process that implies, ideally, no entropy increase of the universe and, therefore, it can be assumed as a  reversible one. In fact, the energy provided by the internal source is kept as `organized' mechanical energy acquired by the system and it can be completely used, at least in principle,  to increase by $-\Delta G_\xi$ the free energy of any chemical reaction.

The next phase of the motion corresponds to the flying period, during which there is no contact with the ground (middle picture in figure \ref{fig:jumper}) and the weight is the only force acting on the person (we are always assuming negligible air resistance forces). However, we are not interested in this part that is not relevant for the thermodynamical description of the overall process.
Instead we want to analyze, from the mechanical and thermodynamical points of view,
the opposite process of the initial part, i.e. the deceleration process of the body when  the person falls down, from the instant he/she touches the ground until stops. This process is represented in part (2) of figure \ref{fig:jumper}. When the feet first comes in contact with the ground (the new initial time), the centre-of-mass is moving downwards with a velocity
$-v_{\rm cm}$   and it stops  after a vertical displacement  $h'$ at a supposedly constant acceleration. The (constant) reaction force is now ${\vec N}'$, pointing upwards, not necessarily equal to ${\vec N}$.
Note that we are assuming the same value for the velocity of the centre-of-mass when the foot abandons the ground, in the ascending trajectory,  and when it comes in contact again with the ground, in the descending trajectory. This means that, in both instants, the centre-of-mass is at exactly the same level with respect to the ground  --- see figure \ref{fig:jumper} --- but, of course, we could relax this restriction.
The dynamics of the descendent body is now described by the integral of (\ref{newtonpt2}) yielding
\beq
 -\left( {1\over 2} M v_{\rm cm}^2 + Mgh'\right) = -N' h'\, .
\label{fg7564}
\eeq
Combining this equation with (\ref{yudfg}) one immediately concludes that
in the ascendent and in the descendent phases the magnitudes of the pseudo-works of the external forces are the same:
\beq
|W'_{\rm ps}|=(N'-Mg)h'= (N-Mg)h = W_{\rm ps}\, .
\eeq
Similarly, denoting by $t_0$ and $t'_0$ the times during which the constant forces $\vec N$ and $\vec N'$ act in phases 1 and 2, the magnitudes of the impulses of the external forces are also the same:
\beq
|I'|=(N'-Mg)t'_0= (N-Mg)t_0 = I\, .
\eeq
Regarding thermodynamics, the situation in the descending phase is not symmetric with respect to the ascendent one. Now there is no internal energy variation (or thermodynamical work and heat) associated with chemical reactions and
equation  (\ref{totale1}), after integration, is now simply given by
\beq
-{1\over 2} M v_{\rm cm}^2 = Mgh' + Q'
\eeq
where $Q'$ denotes the heat exchanged with the surroundings (considered a heat reservoir at temperature $T$). From this equation and from the second equation in (\ref{fg7564})
one obtains \hbox{$Q'=-N'h'<0$}. Again, the normal force doesn't do any work --- it rather intermediates the transformation of mechanical energy into heat. Being a negative quantity, this  is heat transferred from the body to the surroundings.
Considering the overall process --- ascendent and descendent phases ---, and  assuming that the dynamical initial and final states of the body are the same ($h=h'$, then $N=N'$), one concludes that the entropy of the universe
increases by $\Delta S_{\rm U} = {Nh\over T}>0$, due to the heat transfer to the heat reservoir. We note that there is a clear thermodynamical asymmetry between the upward phase and the downward one. Altogether the initial energy provided by the biochemical reactions, producing a decrease
in the Gibbs energy of the body in the first phase of the jump, is totally dissipated in the second phase and not restored, as organized energy, in the person. Therefore one cannot
remain indefinitely jumping vertically in a kind of pendular motion! After some time one needs   something to eat just to keep the jumping activity (of course the basal metabolism also requires nourishment!).

\section{Walking person and other examples}

In the previous  example one part of the process is the time reversal of the other part. Let us now consider an example of a slightly different nature: a walking person.
In a walking person, besides a (time-dependent) horizontal force, there are two vertical forces, namely the
weight and the (time dependent) normal reaction, whose resultant is not exactly zero. In fact, the normal force is, in a certain time interval, larger than the weight and, in another time interval, smaller than
the weight, and this leads to a vertical oscillation
of the centre-of-mass around an average height
with respect to the ground as the person walks \cite{cross99}. In \cite{guemez13c} we discussed the forces on a walking person mainly concentrating on the discussion of the horizontal motion
(a similar discussion may then be applied to the vertical resultant force).

\begin{figure}[htb]
\begin{center}
\hspace*{-0.5cm}
\includegraphics[width=5cm]{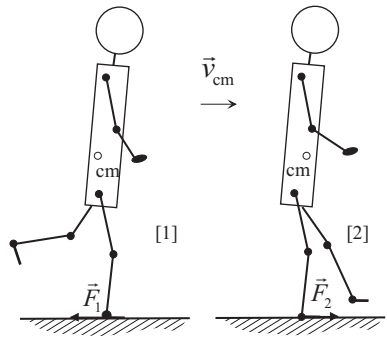}
\end{center}
\vspace*{-0.5cm}
\caption[]{\label{fig:walker} \small the two phases of a stride: (1) the friction force on the foot points backwards; (2) the friction
force on the same foot points forward. }
\end{figure}

Shortly after beginning the hike, a stationary regime is attained in which the speed of the walker becomes (approximately) constant. In fact, the speed of a walking person is not strictly
constant, but there are only small deviations around an average value. The horizontal force
acting upon the  walker's foot is in both the forward and backward directions, in different time
periods that repeat cyclically. The total impulse of that force is zero in a cycle (one cycle is
one stride) so the centre-of-mass linear momentum at the end of the cycle is the same as at the beginning of that cycle.
In a recent paper \cite{haugland13} the horizontal forces acting upon a walking and a running person were presented
(real data) as a function of time but in \cite{guemez13c} we kept the discussion at a simpler level, assuming constant forces in the forward and in the backward directions.
In summary, each stride comprises an accelerating phase, where the centre-of-mass velocity of the body increases (see figure \ref{fig:walker}, part 2), and a decelerating phase were it decreases (part 1 of the same figure).
In the steady state, not only the impulse but also the pseudo
work of the forces (forward, $\vec F_2$, and backward, $\vec F_1$,) vanish, so that after a cycle, there is no variation of the centre-of-mass linear momentum or kinetic energy: walking is a succession
of accelerations and decelerations, but for a cycle
there is no variation of the linear momentum and of the kinetic energy,  $\Delta p_{\rm cm} =0$ and $\Delta K_{\rm cm} =0$.

Let us now discuss the energetic issues.  Walking is possible since the walker possesses an internal source of energy. In the steady regime, the kinetic energy at the beginning is equal to the kinetic energy at the end of the stride.
 In the part of the stride corresponding to the increase of the centre-of-mass kinetic energy, $\Delta K_{\rm cm,2}>0$ and such increase results
 from the biochemical reactions taking place in the person's muscles.
A discussion similar to the one presented in the previous sections leads us to $\Delta K_{\rm cm,2}= - \Delta G_\xi>0$. However, for the other part of the stride, assuming the simplified model used before, the first law of thermodynamics leads
to  $\Delta K_{\rm cm,1}= Q<0$, i.e. some ``organized" energy --- part of the kinetic energy --- is dissipated as heat in the surroundings. Since $\Delta K_{\rm cm,2}=-\Delta K_{\rm cm,1}$ for a stride, globally we can write
for each cycle $\Delta G_\xi=Q<0$. In \cite{guemez13c} we discussed the motion of a toy with an internal source of energy provided by a spring. The conclusion was that $Q=\Delta U$, where $\Delta U<0$ is the
decrease of internal energy. In the present case the internal source of energy is provided by biochemical reactions that take place (in our simple model) at constant temperature and pressure. So the relevant thermodynamical potential is the Gibbs free energy instead of the internal energy. It is again worth noting that all these energy transfers are intermediated by forces that do not perform any work.

In the first part of the stride we may assume that there is no entropy increase: the energy coming from a myriad of microscopic degrees of freedom converges into a single degree of freedom, namely as kinetic energy of the centre-of-mass. However, in the other part of the stride the global kinetic energy does not flow reversibly to each one of the original degrees of freedom. This is the essence of the irreversible character of the
second part of the stride, and therefore of the irreversible character of the whole stride. Irrespective of the model used to describe the walking
process, more or less refined, the following important result never changes:
the free energy (or, if appropriate, the internal energy) variation of the moving body is not recoverable by the body. Instead, it is ``lost" (as organized energy) in the sense that it is dissipated in the surroundings. 

There are many other mechanical systems whose analysis is similar to the one presented in this and in the previous sections. For instance, the boy on rollers pushing against a wall described in \cite{guemez13}  is very similar to the example presented in the previous section.
When the boy on rollers, initially at rest and touching an infinite mass wall starts  pushing against it, he acquires a certain centre-of-mass velocity that is kept constant when the contact with the wall ceases.  Again, this process of mechanical (kinetic) energy increase ideally
does not imply an entropy increase of the universe and,
therefore, it can be regarded as a reversible one. In fact, the kinetic energy of the system can be used completely,
at least in principle, for instance, to increase by $-\Delta G_\xi$ the free energy of a chemical reaction. Let's look now to the time reversal process: the boy on rollers with the same velocity, $v_{\rm cm}$, collides with de wall, placing his hands on the wall and eventually stops.  His initial kinetic energy will be  totally dissipated as heat in the surroundings and there is no chance to recover that energy in a useful ``organized" form,
in particular by restoring the initial internal chemical energy of the first part of the process when he started pushing against the wall. Otherwise he would be  able to swing forever between two parallel walls.

A moving car, studied in    \cite{guemez13b}, a falling cat, a long jump, etc. also provide examples whose thermodynamical analysis
is parallel  to the one described in this article.

\section{Conclusions}
\label{sec:conclusions}

We presented and discussed various mechanical systems to illustrate that pendular motions, in the sense that one part of the motion is ``symmetric" with respect to the other part,
may have different thermodynamical descriptions. We studied three examples, all involving a person performing cyclic motions that are dissipative. Therefore, in order to keep
the motions, a source of energy is required. In all examples, that source is an internal one provided by biochemical reactions taking part in the person's body.
Our model for the transformation of chemical energy into mechanical energy, during part of the motion, is a very simplified one.
When the person contracts the arms, or the legs increasing the mechanical energy
there is a transformation of chemical energy (e.g. via adenosine triphosphate - ATP) into mechanical (sometimes, only kinetic) energy, through internal mechanisms.
When the system increases the mechanical energy due to the internal source, the macroscopic process, only in an idealized way as we adopt here, can be
regarded as a reversible one. At this stage, after transformation of free energy into kinetic energy, one may say that the system still keeps the capacity to produce work
in the sense that the energy is still stored in an organized form. However,
in the other part of the cyclic motion, corresponding to a decrease of kinetic energy, that kinetic energy is directly dissipated as heat and the process is definitely irreversible
(we are assuming no temperature changes).
At the end of each cycle, though the system might recover its initial dynamical state, the entropy of the universe  has already increased.

In conclusion, processes undergone by mechanical systems sometimes also require a thermodynamical detailed analysis in order to be better understood, a point of view adopted
 in the most modern physics textbooks  \cite{chabay11x}.

\end{document}